\documentclass[twocolumn,floatfix,superscriptaddress,a4paper,showpacs,showkeys,nofootinbib,reprint,prc,noeprint]{revtex4-1}
\usepackage{epsfig}
\usepackage{latexsym}
\usepackage[colorlinks=true,linktocpage=true,linkcolor=blue,citecolor=blue,allcolors=blue]{hyperref}
\usepackage{url}
\usepackage[utf8]{inputenc}
\usepackage{enumerate}
\usepackage{color}
\usepackage{xcolor}
\usepackage{graphicx}
\usepackage{xfrac}

\usepackage{amsmath}
\usepackage{amssymb}

\usepackage[english]{babel}
\usepackage{hyperref}
\usepackage{url}
\usepackage{needspace}
\usepackage{float}
\hypersetup{
   colorlinks=true, 
  linktoc=all,     
  linkcolor=blue,  
}

\begin{document}

 \title{Transport simulations with a constrained momentum-dependent Chiral Mean Field EoS at different iso-spin fractions}

\author{Jan Steinheimer}
\affiliation{GSI Helmholtzzentrum f\"ur Schwerionenforschung GmbH, Planckstr. 1, D-64291 Darmstadt, Germany}
\affiliation{Frankfurt Institute for Advanced Studies, Ruth-Moufang-Str. 1,  60438 Frankfurt am Main, Germany}

\author{Marcus Bleicher}
\affiliation{Institut f\"{u}r Theoretische Physik, Goethe-Universit\"{a}t Frankfurt, Max-von-Laue-Str. 1, D-60438 Frankfurt am Main, Germany}
\affiliation{Helmholtz Research Academy Hesse for FAIR (HFHF), GSI Helmholtzzentrum f\"ur Schwerionenforschung GmbH, Campus Frankfurt, Max-von-Laue-Str. 12, 60438 Frankfurt am Main, Germany}


\begin{abstract}
We present a comparison of the UrQMD model using a chiral mean field EoS (CMF) with flow and pion production data in heavy ion collisions at low beam energies $E_{\mathrm{lab}}=0.2-2.0 A$ GeV and varying iso-spin fraction. The CMF model parameters are constrained by known properties of the high density equation of state of QCD at varying iso-spin fractions. This allows us to calculate the equation of state as well as nuclear interactions for different physical systems like neutron stars and heavy ion collisions in a consistent way.
It is found that heavy ion reactions at the upcoming FAIR facility will only have marginal sensitivity on the iso-spin dependence of the high-density equation of state. At lower beam energies, comparing to FOPI, HADES and S$\pi$rit data, the sensitivity is higher but the observed deviations and systematic uncertainties of the existing data are larger than the sensitivity to the symmetry energy.
\end{abstract}

\maketitle
\section{Introduction}

The equation of state of dense QCD matter is being studied extensively in large scale experiments at collider facilities, as well as astrophysical laboratories \cite{Oertel:2016bki,Sorensen:2023zkk}. In particle accelerators one brings heavy nuclei to almost the speed of light before they collide and creates a short lived dense and hot fireball whose remnants can be studied with large detector systems. From the properties of the hadrons emitted from these violent collisions one can infer information on the properties of the compressed matter created which can reach densities up to several times that of normal nuclear matter. The dense matter created in these nuclear collisions is almost iso-spin symmetric, i.e. it consists of similar numbers of protons and neutrons, neutron stars contain another form of dense matter. While here, one also finds densities that reach up to 6 times nuclear saturation density, the iso-spin content of the matter is very asymmetric. The challenge is now to combine these experimental results and draw conclusions on the properties of dense QCD matter in both iso-spin regimes. The goal here is to determine whether, and if so at which density, a transition from a normal nuclear liquid to either a deconfined phase of quarks or a phase of chirally restored matter occurs. 

In neutron stars the equation of state (EoS) can be inferred from measurements of the mass and radius of the star, based on the Tolman-Oppenheimer-Volkoff (TOV) equation \cite{Akmal:1998cf,Lattimer:2006xb}. In heavy ion reactions, due to the non-equilibrium nature of the system created, extensive microscopic transport simulations are necessary to study the EoS \cite{Aichelin:1987ti,Cassing:1990vf,Danielewicz:1999zn,Hartnack:2005tr,Soloveva:2021quj,OmanaKuttan:2022the,Steinheimer:2022gqb}. The challenges in these simulations are manyfold, ranging from how the EoS is implemented in the interactions to the difficulty of describing a relativistic system of interacting particles \cite{Zhao:2025oor}.

The 'standard' observables in heavy ion reaction, to constrain the EoS, are collective flow and its higher order harmonics \cite{Stoecker:1986ci,Aichelin:1987ti,Danielewicz:2002pu,Tsang:2008fd,LeFevre:2015paj,Oliinychenko:2022uvy,OmanaKuttan:2022aml}, as well as particle production near or below their elementary threshold \cite{Aichelin:1987ti,Hong:2013yva,Fuchs:2000kp,Hartnack:2005tr}. Many of the observables and experiments have been thus constraint to lower beam energies and therefore lower baryon densities.
The combination of heavy ion and astrophysical constraints is usually been done by assuming some parametrization of the symmetry energy and then performing a Bayesian inference \cite{Huth:2021bsp}. Here, the results may depend strongly on the choice of observable and beam energy selection. In addition, little is known about the density dependence of the symmetry energy. Suggested probes for the symmetry energy in heavy ion collisions mainly focus on the difference of flow of protons and neutrons \cite{Li:2008gp,Tsang:2008fd,Tsang:2012se} and charged pion production \cite{Li:2002xq,Li:2004cq,SpiRIT:2021gtq} with yet inconclusive results.

In previous works we introduced an approach to this problem by employing a chiral mean field (CMF) model to obtain the EoS which can consistently be applied to heavy ion reactions and neutron star observables \cite{Steinheimer:2024eha}. The connection between these two regimes is based solely on the parameters of the CMF model and can be fixed by input from chiral perturbation theory calculations around saturation density.
In this paper we present results where we extended our UrQMD-CMF simulations to EoS's with different iso-spin per baryon. With this approach we can study how sensitive different observables are on the iso-spin fraction. In addition, we also study the consistency between different experiments, at the same energy, which is relevant to understand the systematic uncertainty in determining the EoS.

\section{The constrained CMF equation of state}

The chiral mean field model (CMF) has been developed to describe the phenomenology of interacting hadrons and quarks from low to high density and temperature. 
It is based on the interactions of baryons and quarks with scalar and vector mean fields which are controlled by a chirally invariant potential. Chiral symmetry is realized in this model by a degeneracy of the ground state baryons with their chiral partners, baryons with the same quantum numbers but opposite parity. If chiral symmetry is restored their masses become equal but stay finite. 

The CMF model in its current form was introduced in \cite{Steinheimer:2011ea,Motornenko:2020yme,Steinheimer:2025hsr} (see \cite{Negreiros:2026ode} for the full Lagrangian and standard set of parameters) also includes a hadron resonance gas of baryons (beyond the octet) and mesons \cite{ParticleDataGroup:2022pth}, which is useful for describing high-temperature physics and is in agreement with lattice QCD results~\cite{Bollweg:2022fqq,Borsanyi:2012cr}.
The Lagrangian ${\cal L}_{{\rm SU}(3)}$ consists of scalar and vector mean-field interactions among the ground-state octet baryons and their parity partners:
\begin{align}
{\cal L}_{\rm{SU(3)}}={\cal L_B}  + U_{\rm scalar} + U_{\rm vector} \,.
\end{align}
 $U_{\rm scalar}$ is the potential of the scalar $\sigma$ and $\zeta$ fields, and $U_{\rm vector}$ is the potential of the vector $\omega$, $\rho$, and $\phi$ fields.

The effective masses of the ground-state octet baryons and their parity partners are
\begin{align}
m^*_{b\pm} &= \sqrt{  (g^{(1)}_{\sigma b} \sigma + g^{(1)}_{\zeta b}  \zeta )^2 + (m_0+n_s m_s)^2 } \pm g^{(2)}_{\sigma b} \sigma\,,
\end{align}
where the indices $(1)$ and $(2)$ refer respectively to the two couplings that determine the common mass and mass difference of the parity partners. The coupling constants $g^{(i)}_{ib}$ are determined by vacuum masses and by nuclear matter properties. $m_0$ refers to a bare mass term of the baryons which is not generated by the breaking of chiral symmetry, and $n_s m_s$ is the ${\rm SU}(3)$-breaking mass term that generates an explicit mass corresponding to the number of strange quarks $n_s$ in a baryon. The values of the mean fields are controlled by the scalar and vector potentials:

\begin{align}
U_{\rm scalar} & =  V_0 - \frac{1}{2} k_0 I_2 + k_1 I_2^2 - k_2 I_4 + k_6 I_6   \nonumber \\
& + k_4 \ln{\frac{\sigma^2\zeta}{\sigma_0^2\zeta_0}} - U_{\rm sb} \,,
\label{veff}
\end{align}
with
\begin{align}
    I_2 &= (\sigma^2+\zeta^2)\,,\nonumber\\
    I_4 &= -(\sigma^4/2+\zeta^4)\,,\nonumber\\
    I_6 &= (\sigma^6 + 4\, \zeta^6)\,,
\end{align} 
\begin{align}
U_{\rm vector}&= -\frac12\left(m_\omega^2\omega^2 + m_\rho^2\rho^2 + m_\phi^2\phi^2\right)\nonumber\\ &-g_4\left(\omega^4+6\beta_2\omega^2 \rho^2+ \rho^4+                 \frac12\phi^4\left(\frac{Z_\phi}{Z_\omega}\right)^2 \right.\nonumber\\
&+3\left.\left(
\rho^2+\omega^2\right)\left(\frac{Z_\phi}{Z_\omega}\right)\phi^2\right)\, ,
\end{align}

where $V_0$ ensures that the pressure in vacuum vanishes. All specific parameters have been provided in \cite{Negreiros:2026ode}.

\begin{figure} [t]
    \centering
    \includegraphics[width=\columnwidth]{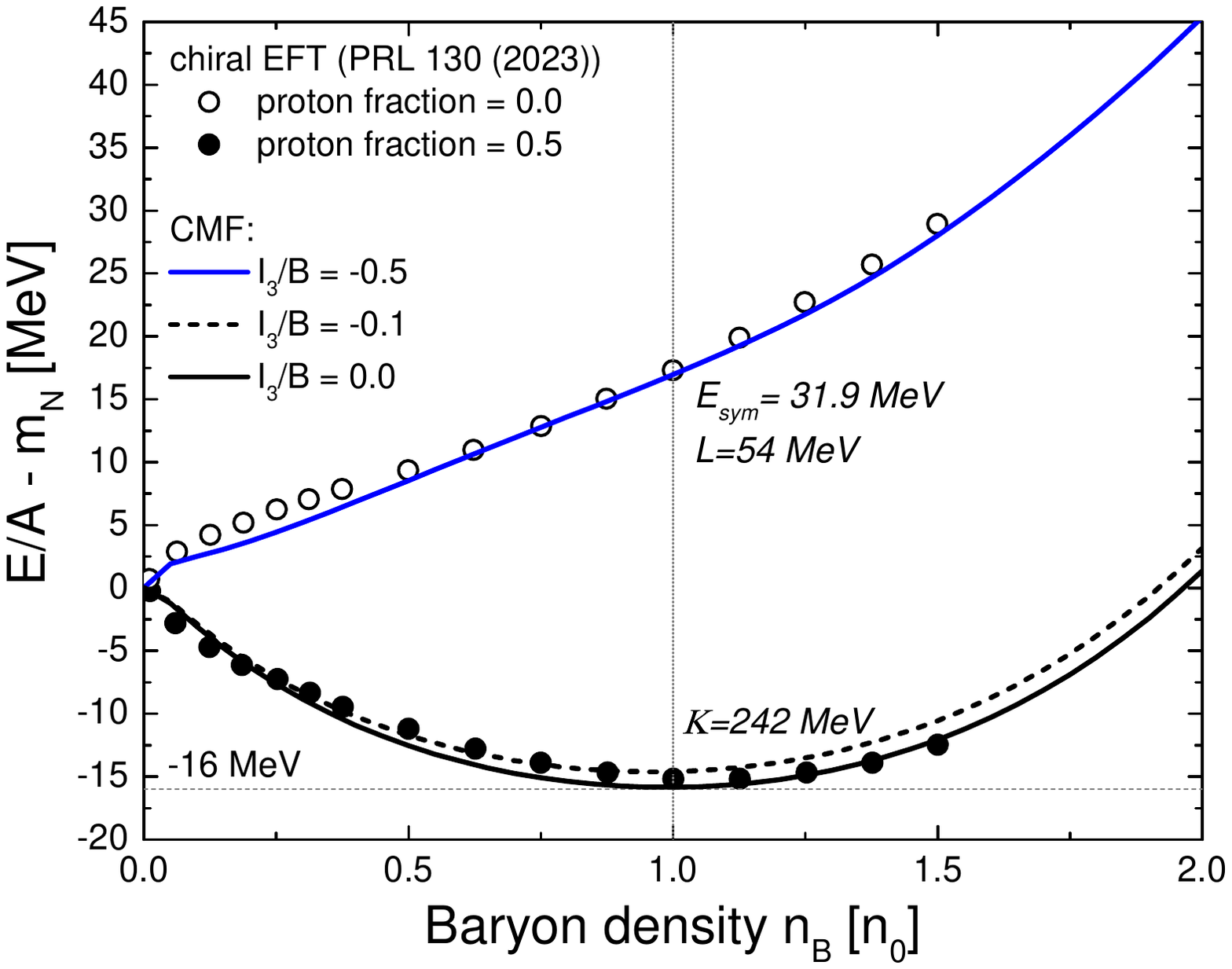}
    \caption{Density dependence of the the nuclear equation of state of iso-spin symmetric matter and pure neutron matter. Compared to chiral EFT results at different iso-spin fraction (filled symbols symmetric matter open symbols pure neutron matter) \cite{Keller:2022crb}. The lines correspond to CMF results where the dashed lines depicts an iso-spin fraction expected for Au nuclei.}
    \label{fig:eosdens}
\end{figure}

In the following we will mainly modify the bare mass $m_0$, two potential parameters $k_0$ and $k_6$ and the scalar and vector couplings $g^{(i)}_{\sigma b}$ and $g^{(i)}_{\omega b}$ to change the nuclear saturation properties while fixing the vacuum masses of the nucleon and its parity partner the $\mathrm{N}^*(1535)$. 
The vector interactions lead to a modification of the effective chemical potentials for the baryons and their parity partners
\begin{equation}
    \mu^*_b=\mu_b-g_{\omega b} \omega-g_{\phi b} \phi-g_{\rho b} \rho \,,
\end{equation}
where the physical chemical potential is $\mu_b$. Again we will mainly modify $g_{\omega b}$ to match the nuclear saturation properties and $g_{\rho b}$ to obtain the correct symmetry energy.

\begin{figure} [t]
    \centering
    \includegraphics[width=\columnwidth]{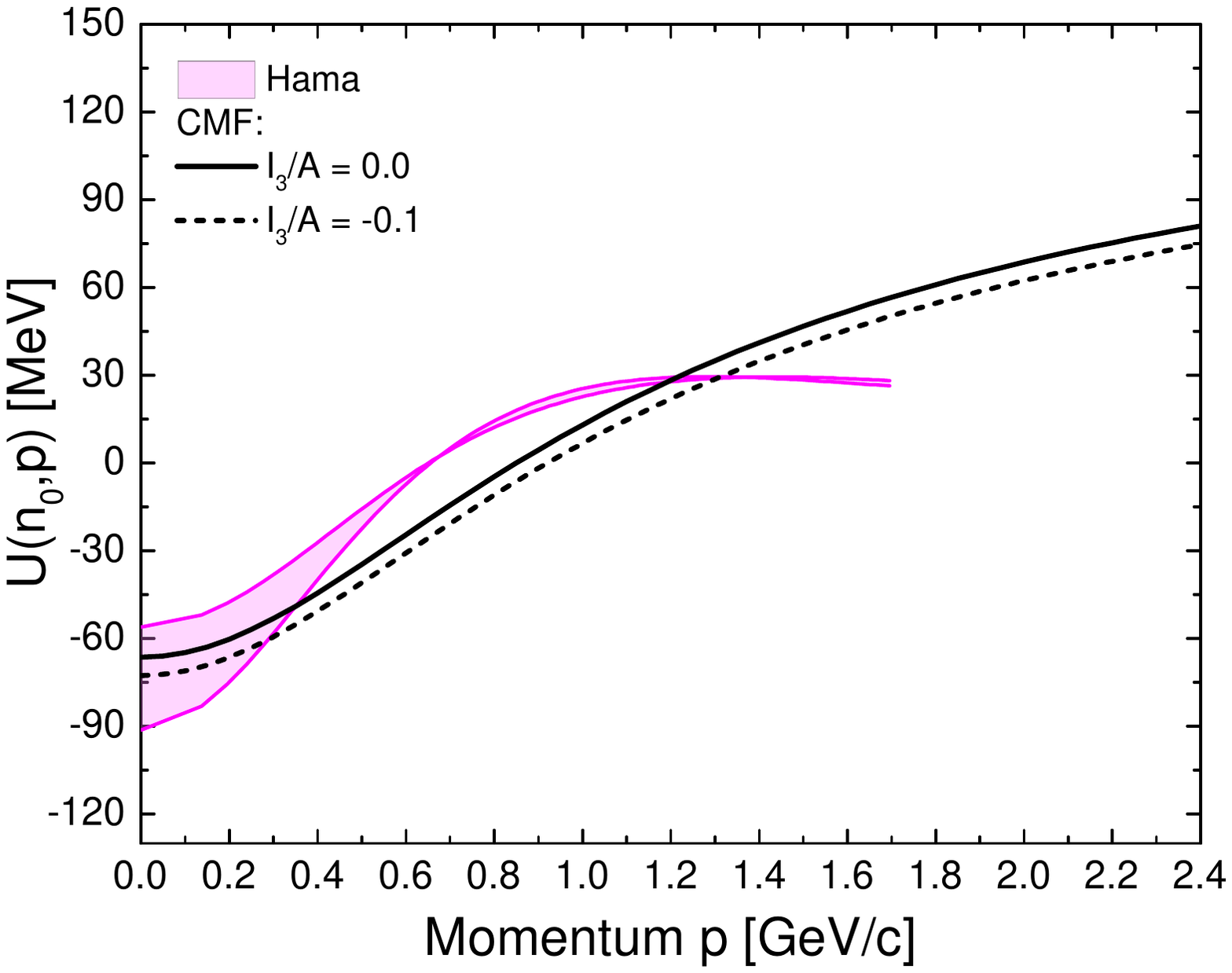}
    \caption{Momentum dependence of single particle energy from CMF compared to two fits to data by Hama et.al. \cite{Hama:1990vr}.}
    \label{fig:hama}
\end{figure}

\subsection{Nuclear saturation properties}

First, we match the model parameters to known properties of nuclear matter. This includes the nuclear saturation density $n_0=0.16 \mathrm{fm^{-3}}$ and binding energy $E/A=-16$ MeV. In addition, the nuclear incompressibility is fixed to $K = 242$ MeV which is also in accordance with expectations \cite{Blaizot:1980tw,Youngblood:2004fe,Garg:2018uam}. To perform calculations for iso-spin asymmetric matter one has to modify the iso-spin chemical potential of the thermal system to obtain the desired iso-spin per baryon $I_3/B$. For symmetric matter this is equal to 0 and for pure neutron matter one has $I_3/B=-0.5$. For matter that is created in heavy ion reactions the lowest iso-spin per baryon is approximately that in Au nuclei: $I_3/B=-0.1$. 
To be able to match the constraints above, one needs to adjust some parameters previously specified in \cite{Negreiros:2026ode}. The modified parameters are given in Table \ref{tab:param}.

Hyperons are fully included in the CMF calculations for all iso-spin fractions, however, when we consider matter in heavy ion collisions, we additionally impose the constraint of strangeness neutrality which essentially removes hyperons at $T=0$.

Figure \ref{fig:eosdens} shows the resulting density dependence of the energy per baryon for iso-spin symmetric matter (black solid line) pure neutron matter (blue solid line) and matter with $I_3/B=-0.1$ (black dashed line), both calculated in the thermal limit for infinite matter at $T=0$~MeV. The CMF results are compared to recent chiral effective field theory (EFT) calculations which should be reliable up to roughly nuclear saturation density. The values for the nuclear symmetry energy $E_{sym}=31.9$ MeV and the slope of the symmetry energy $L=54$ MeV are also indicated in the figure and are consistent with expectations. In general the CMF results are very close to the chiral EFT results. One noteworthy observation is that the energy per baryon for matter created in heavy ion collisions ($I_3/B=-0.1$) is very similar to that of iso-spin symmetric matter.

\begin{table}[b]
\begin{tabular}{|c|c|c|c|c|c|c|}
\hline
 & $g_{\sigma N}^{(1)}$ & $m_0$ & $g_{\omega N}$ & $g_{\rho N}$ & $\lambda_6$ & $k_0 $\\\hline\hline
Value & $-10.78$    & $300$ MeV     & 8.14    & 5.0 & $1.35 10^{-4} \mathrm{MeV}^2$ & $571^2 \mathrm{MeV}^2$\\
\hline
\end{tabular}
\caption{CMF coupling parameter values adjusted to match the constraints discussed in the text.}\label{tab:param}
\end{table}

\begin{figure} [t]
    \centering
    \includegraphics[width=\columnwidth]{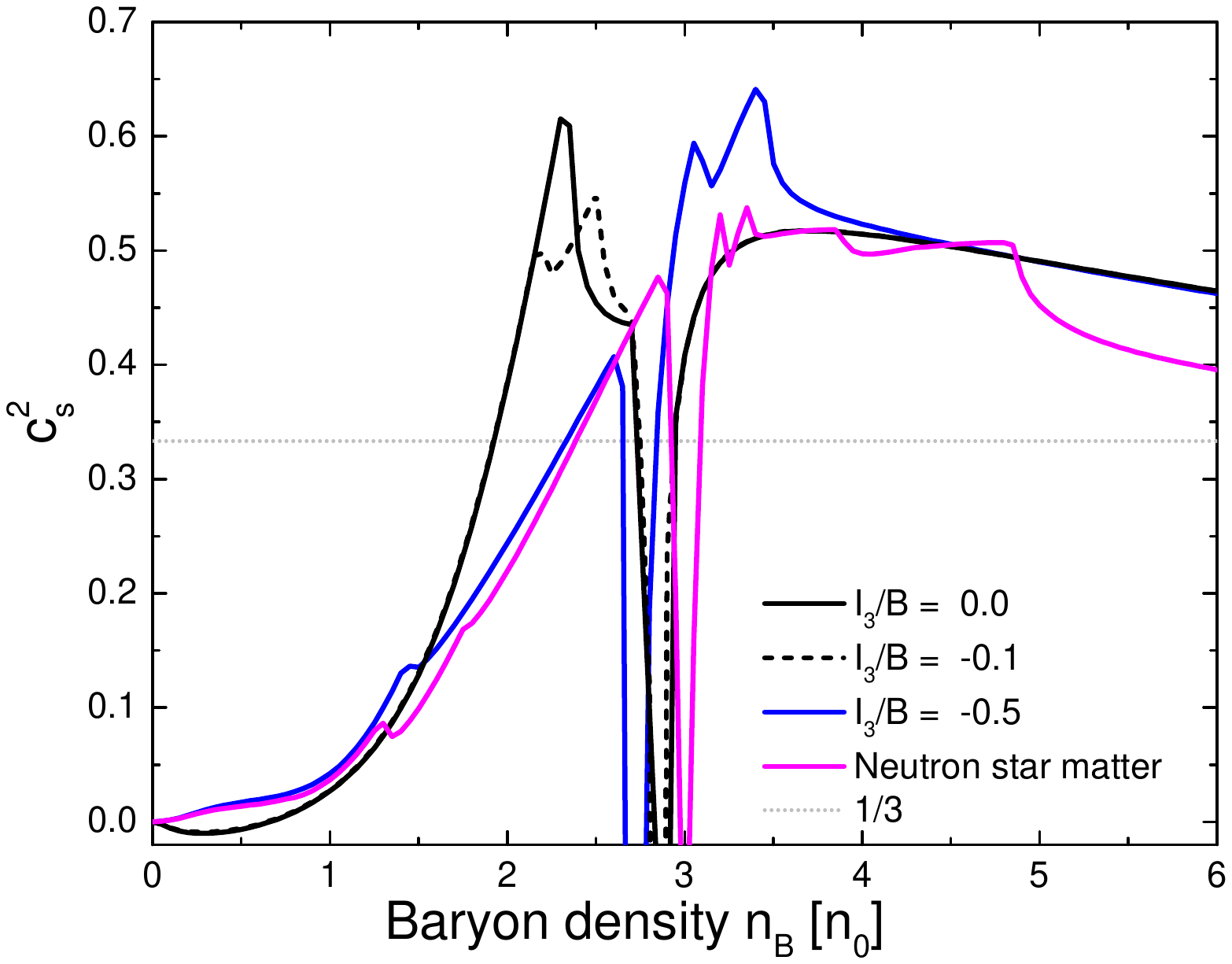}
    \caption{Speed of sound for iso-spin symmetric matter, neutron matter and neutron-star matter, calculated with the same CMF parametrization, as function of density.}
    \label{fig:cs2}
\end{figure}

Let us now turn to the momentum dependence of nuclear potentials. This is an important ingredient to the transport simulations of heavy ion reactions. Several observables, like collective flow and near threshold pion production are sensitive to the momentum dependence of the potential. Constraints on the momentum dependence of the nuclear optical potential are usually based on measurements of the angular distribution of elastic proton-nucleus scattering \cite{Hama:1990vr}, which is fitted within a Dirac potential. The real part of two separate fits of the optical Dirac potential to the scattering is shown as magenta lines in figure \ref{fig:hama}. For comparison we also show the momentum dependence of the mean-field CMF model with our fixed parametrization for iso-spin symmetric matter (solid line) and for heavy ion matter (dashed line), both calculated in the thermal limit for infinite matter at $T=0$ MeV.
While the trend of the fit to the energy dependence of the potential extracted from data by Hama et.al. \cite{Hama:1990vr} is generally described, a small deviation is still observed. In principle one could try to match the parametrization by even better. However, this would lead to an even smaller bare mass term $m_0$ which will inevitably lead to a strong first order phase transition at $\mu_B=0$ which can be excluded by lattice-QCD results. The asymptotic behavior at large momenta is driven by the vector coupling strength which is a constant in the chiral mean field model, while the Hama fit indicates a possible drop towards zero at large momenta. Such a behavior cannot be reproduced within a mean field approach unless the coupling strengths are treated as momentum dependent which is out of the scope of the current work. In general, one can consider these results as the best description of the momentum dependence which can be obtained within the CMF model without violating other constraints. There is a significant advantage of using the CMF for the momentum dependence instead of performing a simple fit (as is usually done), as one naturally obtains the momentum dependence as function of the baryon density in a consistent way.

\subsection{High density behavior}

Direct constraints on the EoS of high density QCD matter come mainly from neutron star properties and neutron star mergers as well as heavy ion reactions. While the latter requires detailed modeling and suffers from systematic uncertainties, the neutron star mass-radius relation provides a more robust constraint, considering both mass and radius are independently observable to high precision. In addition the EoS in neutron stars is different from heavy ion reactions as we can assume the matter to be in $\beta$-equilibrium. That means, neutron star matter includes hyperons. The topic of hyperon potentials is of great interest since they have the potential to connect to neutron star properties (for a recent work see e.g. \cite{Jinno:2023xjr}). In the present paper we do not focus on the effect of the hyperon potentials as we intend a more dedicated work on this in the future.

\begin{figure} [t]
    \centering
    \includegraphics[width=\columnwidth]{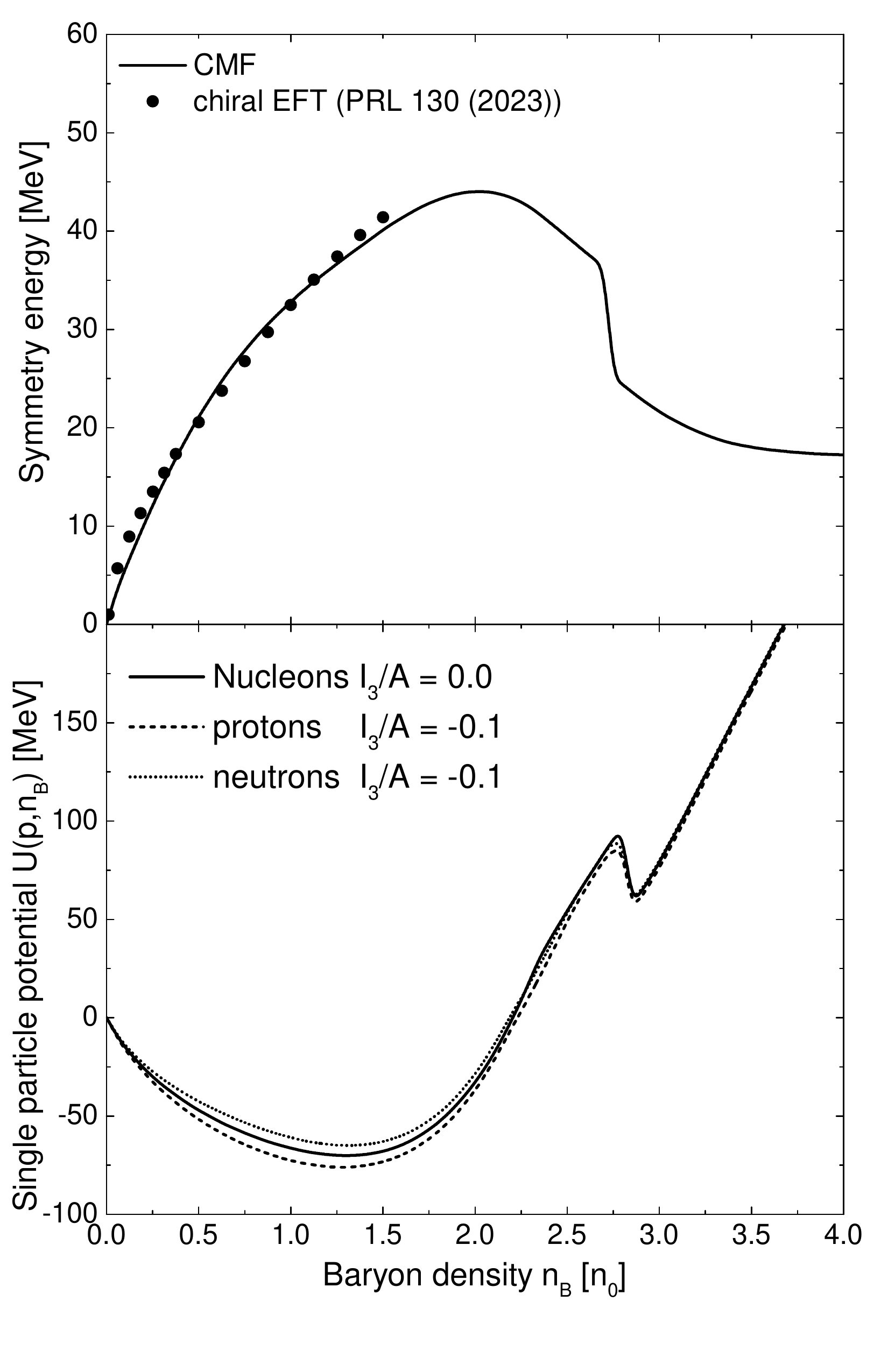}
    \caption{Upper panel: Symmetry energy as function of net baryon density compared to chiral EFT calculations. Lower panel: Momentum dependence of the nucleon potentials in the CMF at $p=0$ for the iso-spin symmetric case (solid line) and for an iso-spin per baryon of $I_3/A=-0.1$ (dashed lines).}
    \label{fig:sym}
\end{figure}

We will first compare the equation of state, represented by the isothermal speed of sound, of the different scenarios in figure \ref{fig:cs2}. In this figure the CMF-EoS for three different values of the iso-spin fraction, but under the assumption of net-strangeness, $s-\overline{s}=0$, is shown. In addition the EoS for neutron-star matter, including $\beta$-equilibrium and leptons is shown as magenta line. All 4 lines show distinct differences. The matter which is close to iso-spin symmetry has a minimum of the speed of sound at low densities, corresponding to bound nuclear matter, but turns stiffer very rapidly. The two strongly iso-spin asymmetric scenarios become softer in the density range between $1.5<n_B/n_0<2.5$. At that point all EoS undergo a chiral phase transition. This transition is rather weak with only a small jump in baryon density. At very high densities the speed of sound will slowly approach the asymptotic limit of an ideal gas of massless quarks. The speed of sound for neutron star matter shows more structure, i.e. more small kinks, which correspond to the appearance of different particles, like hyperons, muons and free quarks.

To better understand the systematics of the symmetry energy in the CMF model, the upper panel of figure \ref{fig:sym} shows the symmetry energy as function of the net baryon density. The CMF results are compared to the chiral EFT results. At low densities, the symmetry energy increases as expected as the Fermi energy of neutrons is increased. At roughly twice nuclear saturation density, the trend is inverted and the symmetry energy starts to decrease and even undergoes a steep drop around the chiral phase transition. This is because the appearance of the parity partner of the nucleon, as well as other degrees of freedom, allow for the distribution of iso-spin over more d.o.f. decreasing effectively the Fermi energy of neutrons. This is an interesting and important results, as it means that the nuclear potential at higher densities will be much less sensitive to the iso-spin content of the system making it more difficult to obtain constraints.

This effect can be observed in the density dependence of the nuclear potential which is shown in the lower panel of figure \ref{fig:sym}, for zero momentum. Here, the solid line depicts the nucleon potential, as $U_{proton}=U_{neutron}$ for iso-spin symmetric matter. The dashed lines correspond to the potentials for protons and neutrons for matter with a small iso-spin asymmetry, as expected for heavy ion reactions. It is clear that the difference between the potentials is largest around nuclear saturation density and then is reduced. At densities above twice nuclear saturation density we observe almost no difference between the two scenarios.

\begin{figure} [t]
    \centering
    \includegraphics[width=\columnwidth]{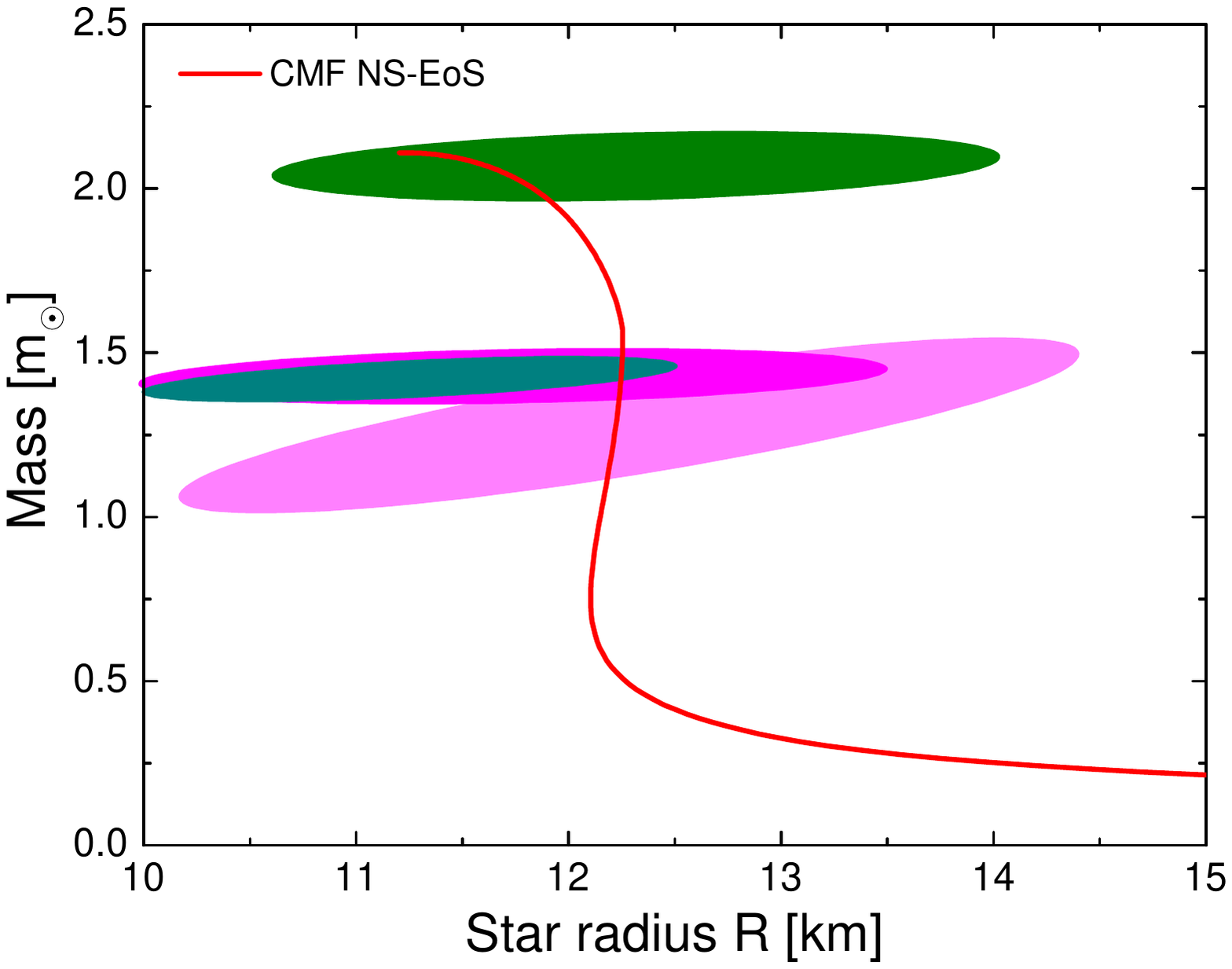}
    \caption{Mass radius relation for neutron star matter from CMF (red line) compared to experimental constraints taken from simultaneous mass-radius measurements \cite{Riley:2019yda,Riley:2021pdl,Choudhury:2024xbk}.}
    \label{fig:mr}
\end{figure}

\subsection{Neutron star properties}

An important constraint on the high density behavior of the QCD EoS comes from the properties of neutron stars and their mergers. For cold, non-rotating, neutron stars the relation of the stars mass and radius can be derived from the Tolmann-Oppenheimer-Volkoff (TOV) equation \cite{Oppenheimer:1939ne}. Using the neutron star EoS calculated with the CMF model we can simply solve the TOV equation for  stars of different initial central density and the results are depicted in figure \ref{fig:mr}. For comparison we also show the constraints from the NICER experiment \cite{Riley:2019yda,Riley:2021pdl,Choudhury:2024xbk}. The results obtained with the CMF model are consistent with all NICER constraints as well as the maximum mass constraint \cite{Demorest:2010bx,Antoniadis:2013pzd,Bauswein:2017vtn,Margalit:2017dij,Rezzolla:2017aly}.

\subsection{Finite temperature}

Even though the potential that is included in the UrQMD simulations does not explicitly depend on Temperature, it is worthwhile to check whether the temperature dependent EoS can be captured within the CMF model. A significant part of the temperature dependence comes from the appearance of pions and other light mesons, as well as baryonic resonance excitation. At much higher temperatures, the contributions of deconfined quarks as well as gluons will start to dominate. All these effects are included in the CMF model. The contribution of the hadronic resonances is simply added to the partition function via a HRG contribution with vacuum masses for all hadrons which do not explicitly couple to the chiral fields. The thermal quark contribution to the thermodynamic potential is:

\begin{equation}
	\frac{\Omega}{V}=-T \sum_q\frac{d_q}{N_c}\int\!\frac{d^3k}{(2 \pi)^3}\ln\left(1+ \Phi\,e^{{-\left(E_q^*-\mu^*_q\right)/T}}\right)\,,
	\label{eq:q}
\end{equation}
with index $q$ running through $u,d,s$ quark flavors, $N_c$ is the number of colors, and $d_{q}$ the color and spin degeneracy factor. The thermal contribution of the gluons is linked to the Polyakov-loop order parameter $\Phi$ and is controlled by the temperature dependent potential~\cite{Motornenko:2019arp}:

\begin{eqnarray}
    U_{\rm Pol}(\Phi,\overline{\Phi},T) &=& -\frac12 a(T)\Phi\overline{\Phi} \\ \nonumber
	 + b(T)\ln \Bigl[1-6\Phi\overline{\Phi}\Bigr.
	 &+& \Bigl. 4(\Phi^3+\overline{\Phi}^{3})-3(\Phi\overline{\Phi})^2\Bigr] \,, \\
    a(T) &=& a_0 T^4+a_1 T_0 T^3+a_2 T_0^2 T^2,  \nonumber \\
    b(T) &=& b_3 T_0^4  \nonumber\,.
\end{eqnarray}

\begin{figure} [t]
    \centering
    \includegraphics[width=\columnwidth]{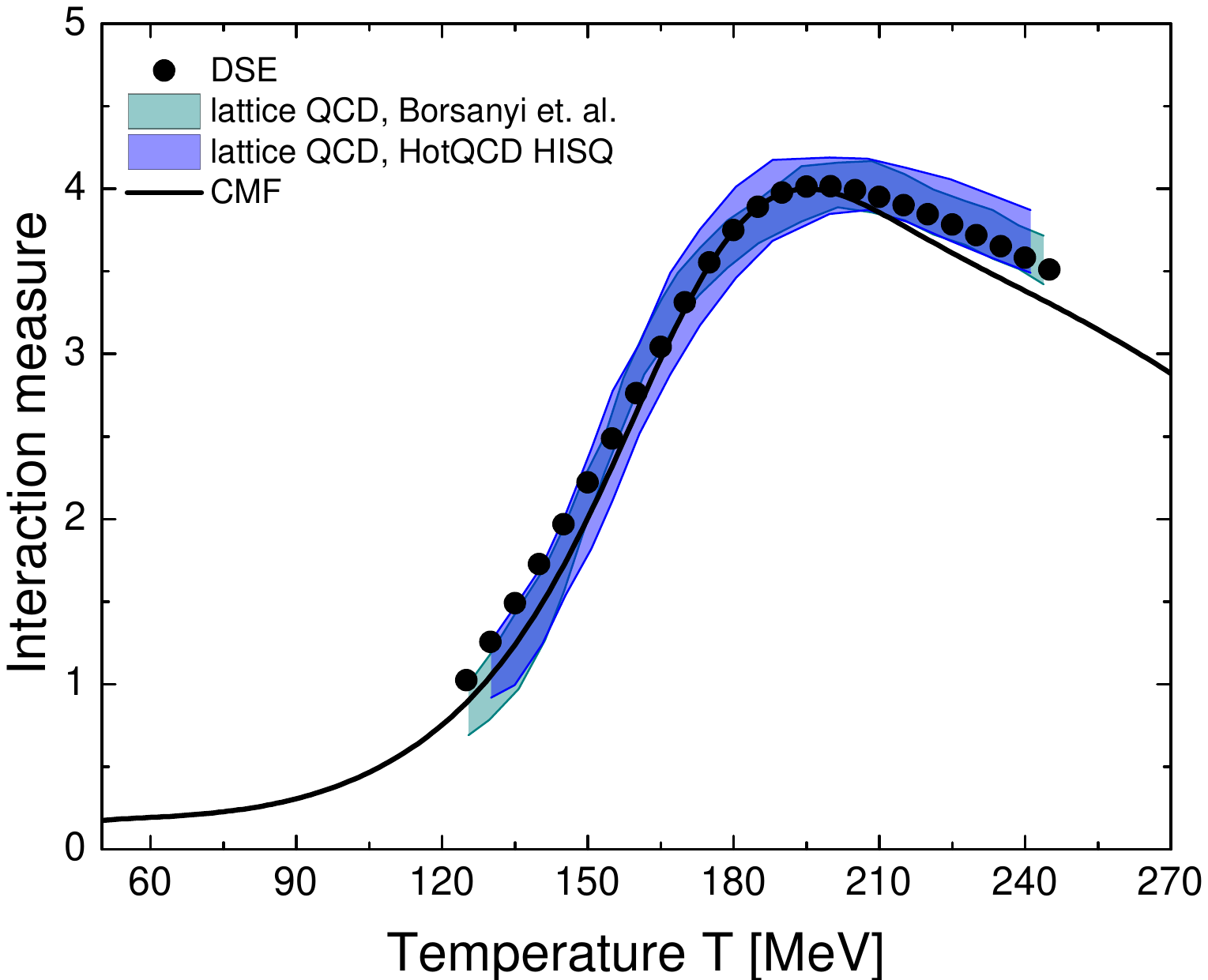}
    \caption{Eos at finite temperature and vanishing net density compared to lattice-QCD data \cite{Bollweg:2022fqq,Borsanyi:2012cr} and a DSE approach to QCD \cite{Lu:2025cls}.}
    \label{fig:lattice}
\end{figure}

The parameters of this potential are fixed by demanding a reasonable description of lattice QCD results at vanishing net baryon density. This comparison is shown in figure \ref{fig:lattice}. For reference we also show the results based on the Dyson-Schwinger approach  \cite{Lu:2025cls}. As one can see the CMF model provides a good description of lattice QCD thermodynamics, especially at lower temperatures where the system consists of confined hadrons. We therefore can safely assume that also the temperature dependence, at least until deconfinement effects occur, in the UrQMD+CMF simulations will be realistically described.

\section{UrQMD implementation}

The implementation of the CMF potential in UrQMD has been described in \cite{Steinheimer:2024eha,Steinheimer:2025hsr}. Here, we want to provide only a short summary. The equations of motion which determine the change of the baryons momentum as well as the velocity with which a particle propagates in a given time-step follow from a Hamiltonian \cite{Aichelin:1986wa}:

\begin{eqnarray}\label{motion}
\dot{\textbf{r}}_{i}=\frac{\partial \mathrm{\bf{H}}  }{\partial\textbf{p}_{i}},
\quad \dot{\textbf{p}}_{i}=-\frac{\partial \mathrm{\bf{H}} }{\partial \textbf{r}_{i}} .
\end{eqnarray}
where $ \mathrm{\bf{H}} = \sum_i H_i$ is the total Hamiltonian function of the system which is the sum over all Hamiltonians, $H_i=E^{\mathrm{kin}}_i + V_i$, of the $i$ baryons. In this Hamiltonian, the kinetic term is the relativistic kinetic energy and the potential term is a non-relativistic density and momentum dependent potential. 

This means that, in the absence of any potential, the correct relativistic velocity follows as $\dot{\textbf{r}}_{i}=\textbf{p}_{i}/E$. In the setup including potentials, the momentum dependence can be understood as mimicking the effect of an effective mass in the relativistic kinetic energy. The change of momentum of each baryon is then calculated from the derivative of the potential energy assuming that each particle can be treated as a Gaussian wave packet~\cite{Aichelin:1991xy,Bass:1998ca}.

The CMF model provides the density and momentum dependence of different baryon (and anti-baryon) potentials, including the lowest baryon octet (with hyperon potentials fitted to describe the known binding energy of hypernuclei) and all its parity partners plus the $\Delta-$baryon in tabulated form \footnote{Let us note that we use the potentials calculated from CMF at vanishing temperature for both iso-spin scenarios. As we discussed before the equation of state itself will depend on temperature due to the thermal contributions of the hadrons even if they do not interact. An explicit temperature dependence of the potentials could be incorporated by introducing potentials which depend on the vector as well as scalar density. This is work that will be presented in a future publication.}. In addition, the effective mass of the baryons as function of density is also tabulated and then used in the calculation of the relativistic velocity in eq.(\ref{motion}). To compare different scenarios of the iso-spin asymmetry and thus quantify the possible affect of the symmetry at large density we can calculate the CMF tables for two fixed values of iso-spin per baryon: $I_3/A=0$ for symmetric matter and $I_3/A=-0.1$ corresponding to the asymmetry found in Au-nuclei.

\begin{figure} [t]
    \centering
    \includegraphics[width=\columnwidth]{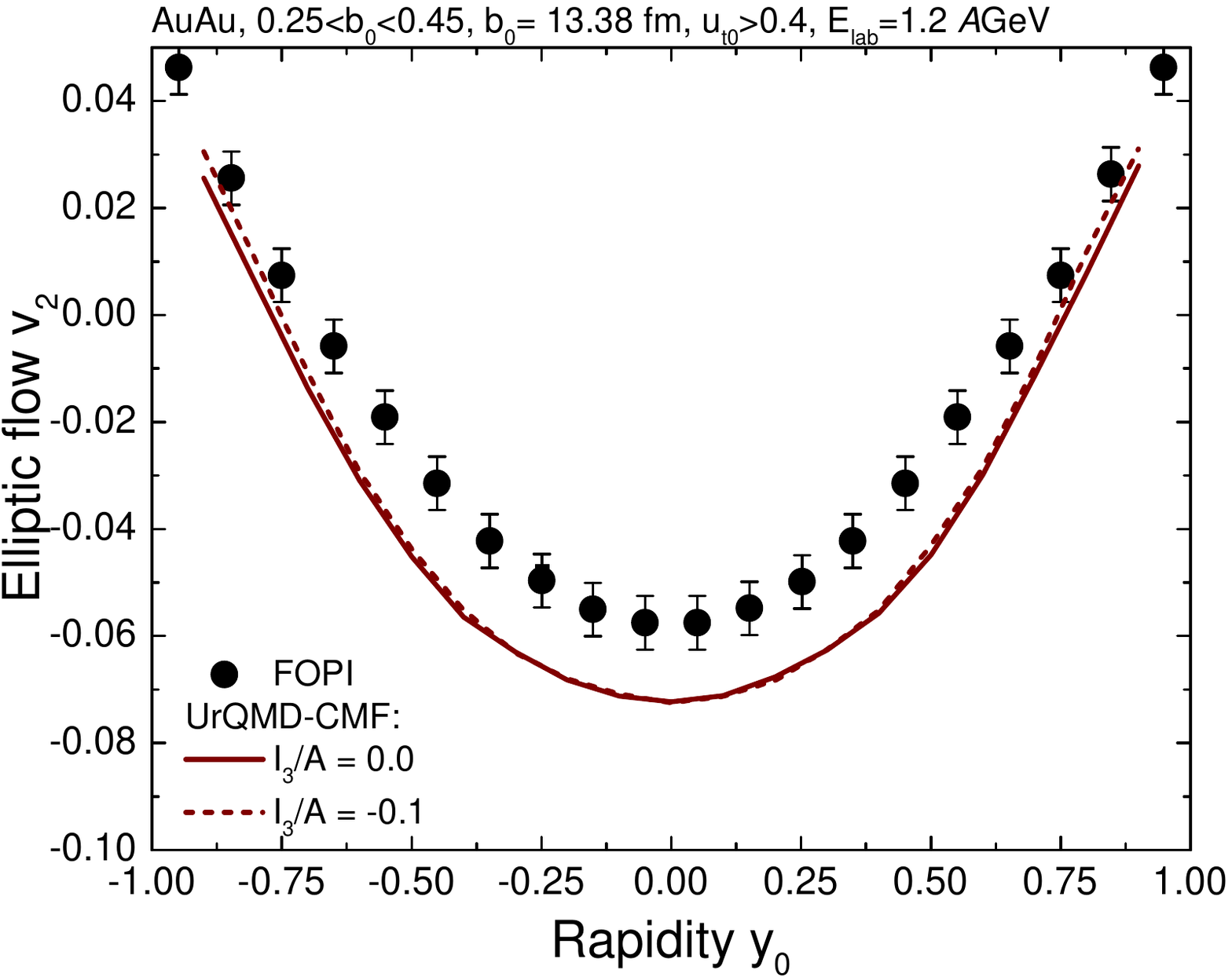}
    \caption{Elliptic flow of protons as function of rapidity for $E_{lab}=1.2 A$GeV. UrQMD-CMF simulations (lines) are compared to data from FOPI \cite{FOPI:2011aa}.}
    \label{fig:v2_y_fopi}
\end{figure}

\begin{figure} [t]
    \centering
    \includegraphics[width=\columnwidth]{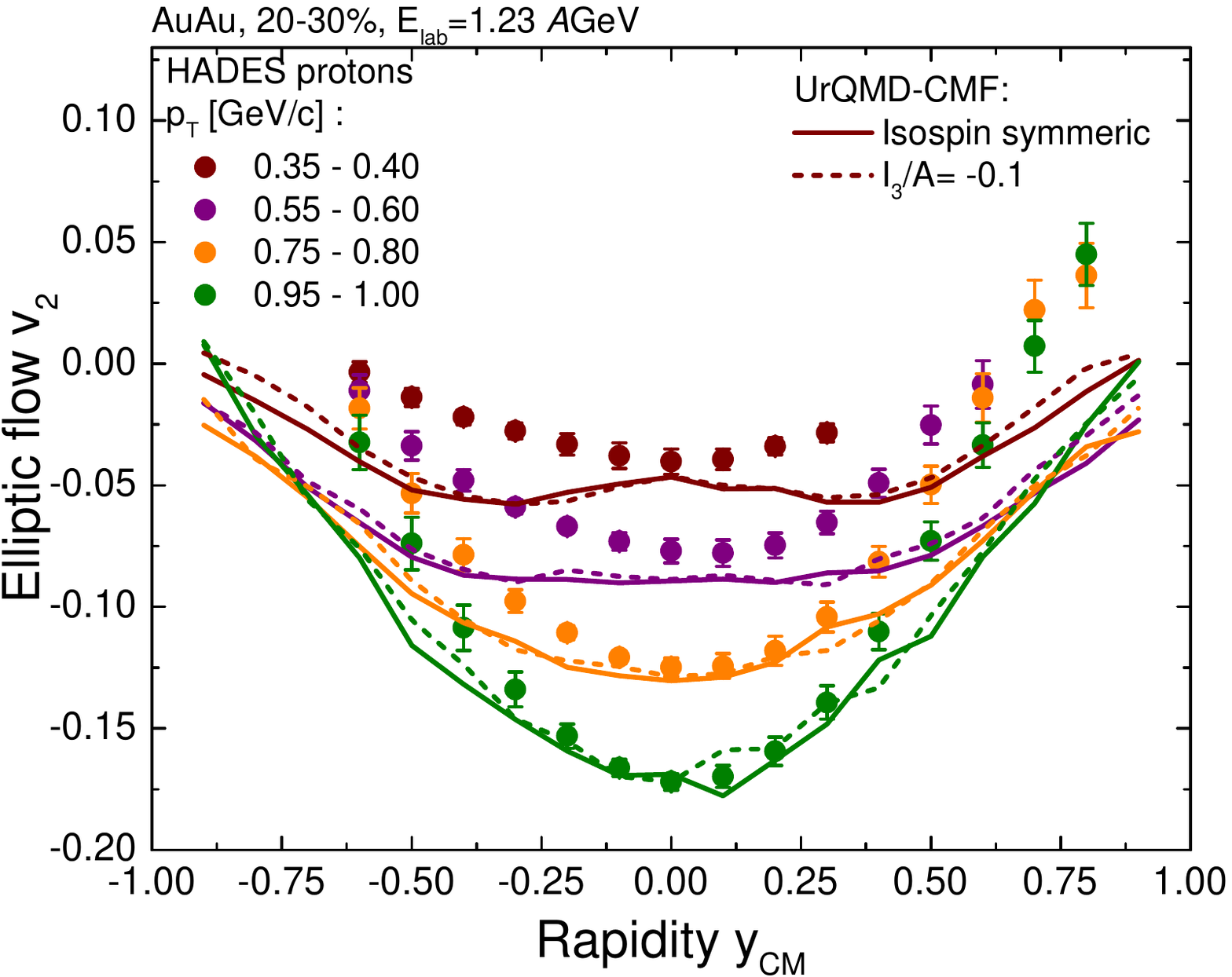}
    \caption{Elliptic flow of protons as function of rapidity for $E_{lab}=1.23 A$GeV. The different colors correspond to different transverse momentum ranges. UrQMD-CMF simulations (lines) are compared to data from HADES \cite{HADES:2020lob,HADES:2022osk}.}
    \label{fig:v2_y_had}
\end{figure}

\section{Results for heavy ion reactions}

Relativistic heavy ion collisions have been used in the past to study the properties of dense and hot QCD matter. Large scale future projects, e.g. at FAIR, aim at studying highly compressed baryonic matter. In the following we will focus on several observables which have been discussed to be sensitive to the density, momentum- and iso-spin-dependence of the nuclear interactions and thus the EoS. The first is the elliptic flow, quantifying the asymmetry of particle emission in the transverse plane. Here, we simply calculate the elliptic flow with respect to the reaction plane defined by the model simulation:

\begin{equation}
v_2= \left\langle\frac{p_x^2 -p_y^2}{p_x^2 + p_y^2}\right\rangle ,
\end{equation}
where the average runs over all particles of a given species. The elliptic flow is usually measured either as function of the transverse momentum or integrated over a certain momentum range and has been shown to be sensitive to the EoS and the momentum dependence of the potentials. In figure \ref{fig:v2_y_fopi} we compare the rapidity dependence of the $p_T$ integrated $v_2$ for two different scenarios of the iso-spin per baryon (solid and dashed lines) with experimental data from the FOPI collaboration. The data where taken in Au+Au reactions at $E_{\mathrm{lab}}=1.2 A$ GeV. The centrality of FOPI is defined with respect to a maximum impact parameter which is different from other heavy ion experiments which define centrality bins in percentiles of the total cross section. 

For reference, the shown data would correspond to mid-central ($\approx 5-15 \%$) collisions. The simulations are performed using the UrQMD-CMF model as described above and using the two iso-spin scenarios. To avoid late stage artifacts, the runtime of the simulations is limited to $t=40 fm/c$. To allow for a realistic comparison to the experimental data, we perform a coalescence calculations to create light nuclear clusters up to $\mathrm{He}$ to obtain the flow of the free nucleons. As we can see, the model simulation gives a reasonable description of the data. The elliptic flow from the model is slightly too large as compared to the data but the centrality dependence and width are reasonably reproduced.

\begin{figure} [t]
    \centering
    \includegraphics[width=\columnwidth]{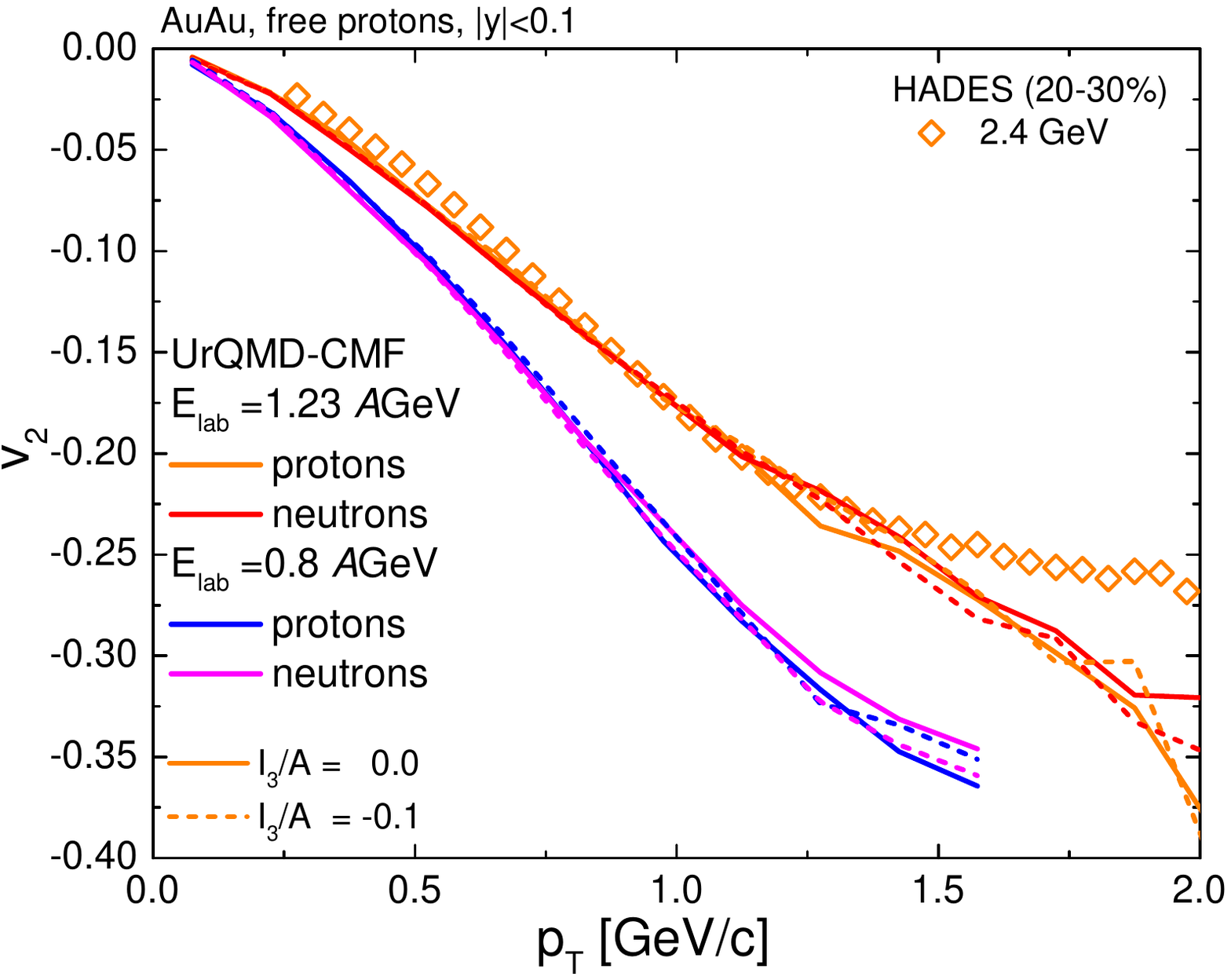}
    \caption{Elliptic flow as function of transverse momentum for mid-rapidity protons. The lines correspond to UrQMD-CMF results for protons and deuterons at two different beam energies and two iso-spin fractions. The symbols correspond to HADES data at $E_{lab}=1.23 A$GeV \cite{HADES:2020lob,HADES:2022osk}.}
    \label{fig:v2_hades}
\end{figure}

To put this comparison into perspective, we can also compare our model results with measurements of the elliptic flow of a different experiment. HADES has measured proton flow in Au+Au collisions at $E_{\mathrm{lab}}=1.23 A$ GeV, which is almost identical with the FOPI measurement. Figure \ref{fig:v2_y_had} shows the rapidity dependence of the elliptic flow, integrated over different transverse momentum intervals. Again, we compare to UrQMD-CMF simulations with two different iso-spin fractions. These results are obtained in a $20-30 \%$ centrality bin, i.e. more peripheral than the FOPI data. However, we can already see a difference. While in the comparison with the FOPI data, the model showed an increase of the magnitude of $v_2$ with respect to the data at mid-rapidity, this is not the case for the HADES data. On the other hand we observe a clear deviation at low transverse momenta and large rapidities, where the contributions from the fragmenting spectator in the UrQMD-CMF simulation cannot be precisely captured. 

\begin{figure} [t]
    \centering
    \includegraphics[width=\columnwidth]{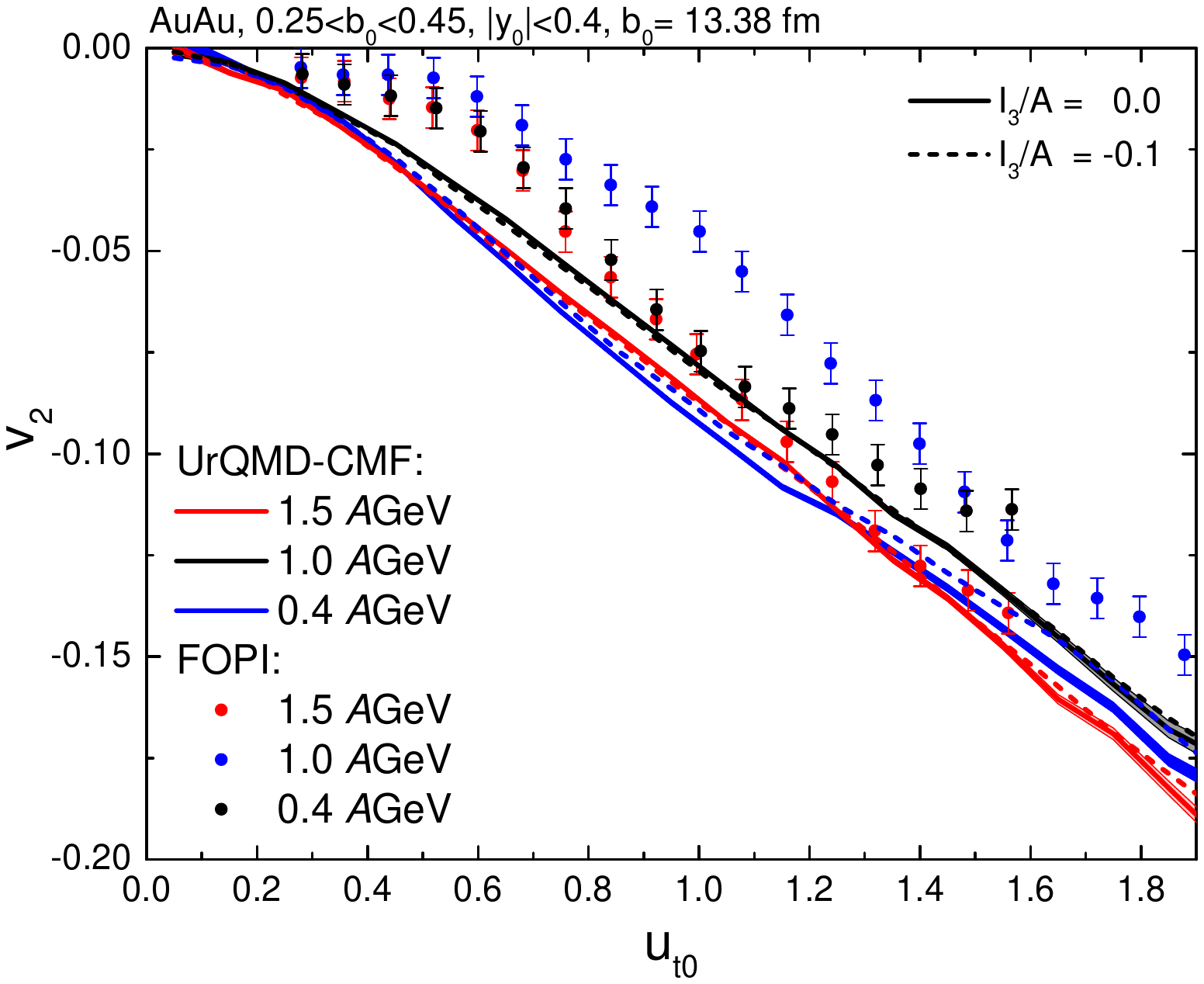}
    \caption{Elliptic flow of protons as function of the scaled transverse momentum (as defined in \cite{FOPI:2011aa}) for mid-rapidity $|y_0| <  0.4$. Shown are three different beam energies where the UrQMD-CMF simulations (lines) are compared to FOPI data.}
    \label{fig:v2_fopi}
\end{figure}

A discrepancy in the model-data comparison can also be observed for the differential $v_2$ measurements. Figures \ref{fig:v2_hades} and \ref{fig:v2_fopi} show the elliptic flow of free protons as function of the transverse momentum for Au+Au collisions at different beam energies. Figure \ref{fig:v2_hades} compares the UrQMD-CMF results with HADES data and \ref{fig:v2_fopi} with FOPI data. Again, both experiments have different rapidity acceptances as well as centrality definitions and selections which makes a direct comparison difficult. However, we can again see that the UrQMD-CMF model compares more favorably with the HADES data (very close to mid-y) than with the FOPI data which has a larger rapidity acceptance. Especially at lower beam energies we see a systematic deviation between model and data at low $p_T$ which is again due to the imperfect description of the fragmenting spectator remnants in the URQMD-CMF model.

Regarding the effect of the iso-spin fraction of the observable flow, all results show that the effect is very small and almost negligible. Figure \ref{fig:v2_hades} even shows the difference between proton and neutron elliptic flow (as different colors) for both iso-spin symmetry scenarios. The difference between the two iso-spin states at the beam energy of $E_{\mathrm{lab}}=1.23 A$ GeV is not visible and at the lower beam energy of $E_{\mathrm{lab}}=0.8 A$ GeV is very small. It is therefore unrealistic to try to constrain the iso-spin dependence of the EoS at high density with elliptic flow measurements in heavy ion reactions in the GeV energy regime.

\subsection{Charged pion production}

The production of charged pions at energies close to threshold has been shown to be sensitive to the momentum dependence of the nuclear potential \cite{Aichelin:1987ti,Hong:2013yva}. It has also been suggested that the ratio of negative to positive charged pions may carry some information on the iso-spin dependence of the interactions \cite{Li:2002xq,SpiRIT:2021gtq,Li:2004cq}. In the following we will compare results with the new UrQMD-CMF model and two different iso-spin symmetries with experimental results on pion production from the FOPI, HADES and S$\pi$RIT experiments. To be able to do so we have to make certain improvements to the UrQMD model. In its current version, UrQMDv4.0 does not include any Coulomb interactions for charged mesons. Since the Coulomb interaction will be important for the difference in the transverse momentum distributions, we have now implemented the Coulomb interactions also between mesons and mesons as well as mesons and baryons.

\begin{figure} [t]
    \centering
    \includegraphics[width=\columnwidth]{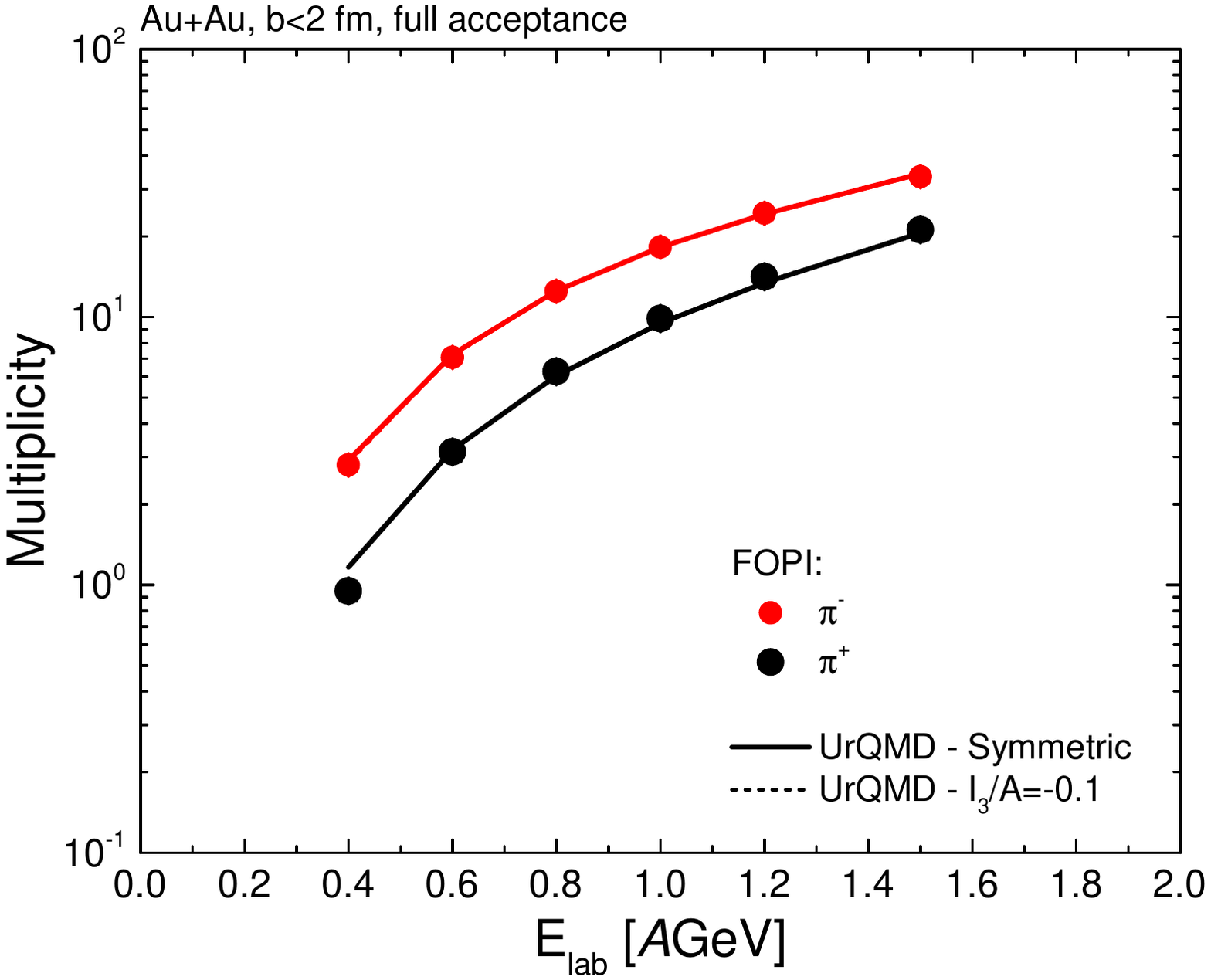}
    \caption{Charged pion multiplicities for central AuAu collisions from the UrQMD-CMF model with different iso-spin fractions (solid and dashed blakc lines), compared to FOPI data \cite{FOPI:2006ifg}.}
    \label{fig:pi_fopi}
\end{figure}

Figure \ref{fig:pi_fopi} shows the beam energy dependence of the mean charged pion numbers for central Au+Au collisions for the two iso-spin symmetry scenarios. The UrQMD-CMF results are compared to FOPI data. No significant difference between the data nad the model for the charged pion production is observed. The model simulations give a good description of the data, except for the lowest beam energy, where the $\pi^+$ production is slightly overestimated in the model.

\begin{figure} [t]
    \centering
    \includegraphics[width=\columnwidth]{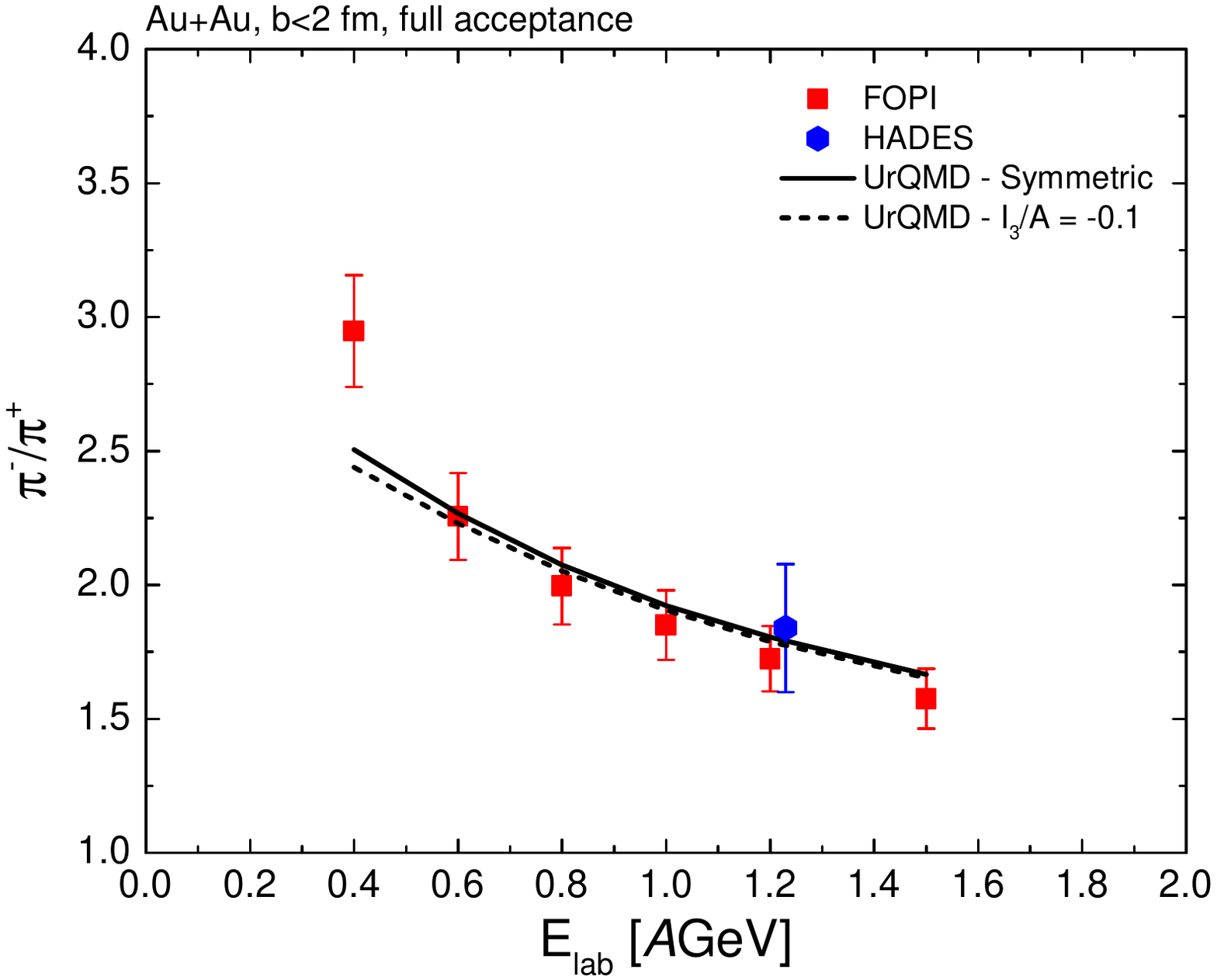}
    \caption{Charged pion ratios for central AuAu collisions compared to FOPI \cite{FOPI:2006ifg} and HADES \cite{HADES:2020ver} data.}
    \label{fig:pi_ratio_fopi}
\end{figure}

This difference can be more clearly observed when the ratio of negative to positively charged pions is taken. This ratio is shown as function of beam energy in figure \ref{fig:pi_ratio_fopi} and compared to FOPI and HADES data. Again, all beam energies except the lowest are well described and only a very small effect from the iso-spin asymmetric potentials is observed.

\begin{figure} [t]
    \centering
    \includegraphics[width=\columnwidth]{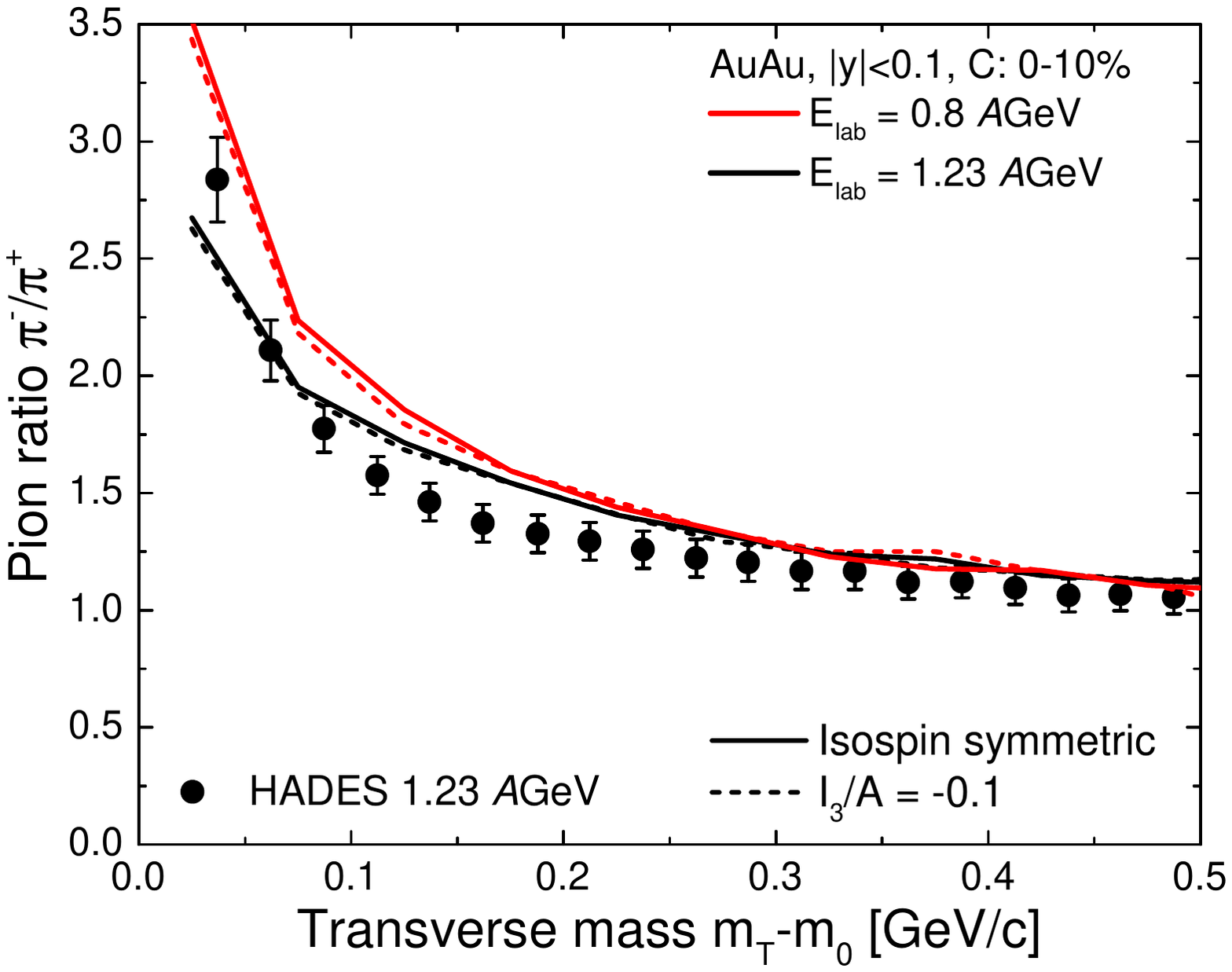}
    \caption{Charged pion ratio as function of mean transverse mass for collisions of AuAu $E_{lab}=1.23$ and $0.8A$GeV. Symbols correspond to HADES data at $E_{lab}=1.23$ \cite{HADES:2020ver}}
    \label{fig:pi_ratio_had}
\end{figure}

When looking at the transverse mass dependence of the charged pion ratio, the effect of the Coulomb interaction becomes dominant. This can be seen in figure \ref{fig:pi_ratio_had} where the UrQMD-CMF simulations for the two iso-spin fraction cases are compared to HADES data at $E_{\mathrm{lab}}=1.23 A$ GeV (black lines and symbols). We also present a prediction for the ratio at a lower beam energy of $E_{\mathrm{lab}}=0.8 A$ GeV where the effect becomes stronger. We can observe again, that the data is well described by both iso-spin asymmetry scenarios with only a very small effect from the finite iso-spin density. The main effect at low $p_T$ is given by the Coulomb interaction.


\subsubsection{S$\pi$RIT}

The S$\pi$RIT experiment has performed measurements of pion production at beam energies very close to the production threshold and for systems of varying iso-spin asymmetry. In particular three systems have been investigated: Sn(108)+Sn(112), Sn(112)+Sn(124) and Sn(132)+Sn(124) at a beam energy of $E_{\mathrm{lab}}=0.27 A$ GeV. The centrality selection for these systems corresponds to an impact parameter of $b=3$ fm and we will show results for polar angles, in the center-of-mass frame of $\theta<90°$. The different collision systems have a neutron over proton fractions $N/Z$ which can also be related to the iso-spin fraction. Note, that the largest $N/Z$ investigated in these studies corresponds to an iso-spin fraction which is just marginally different from that in Au+Au. 

\begin{figure} [b]
    \centering
    \includegraphics[width=\columnwidth]{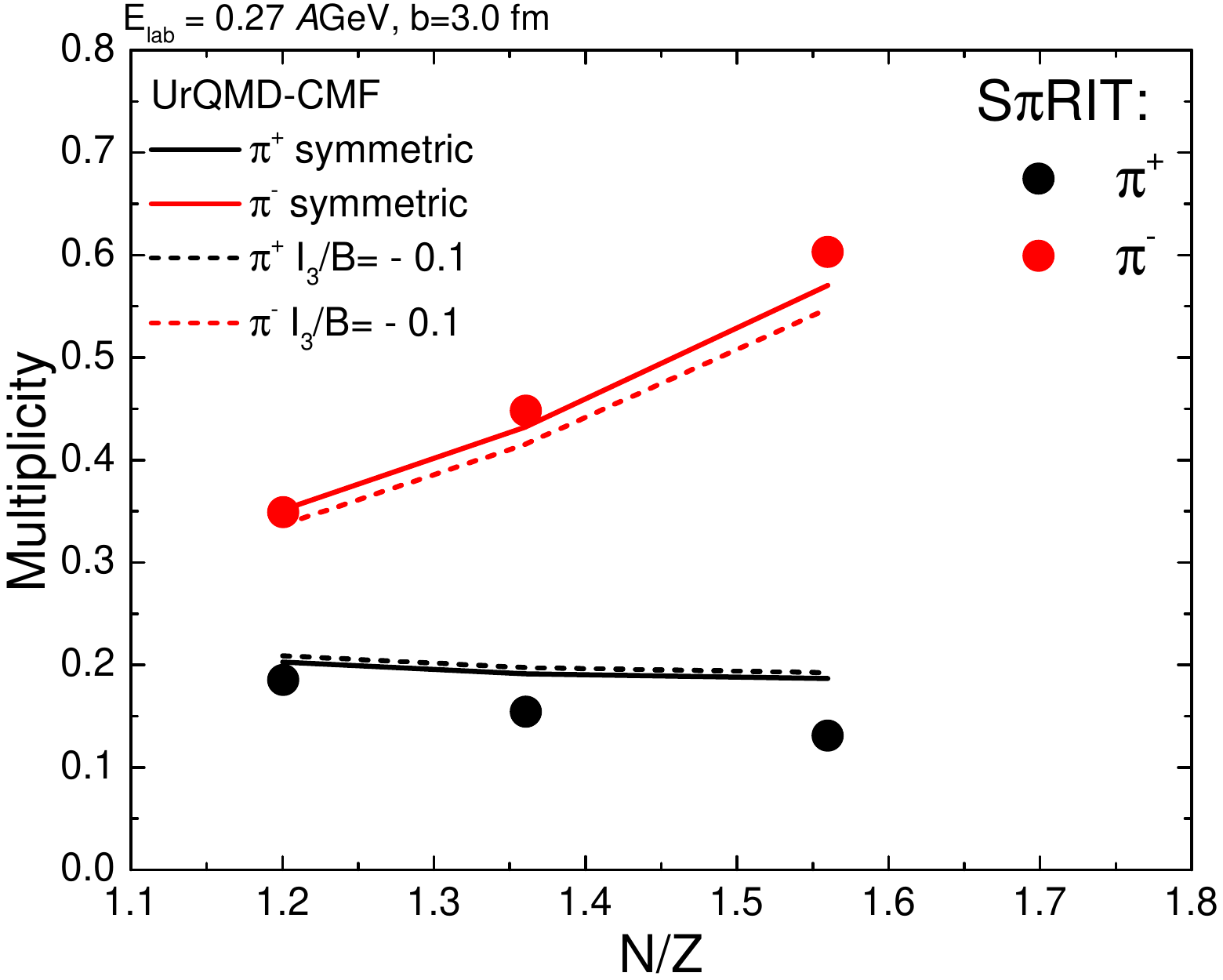}
    \caption{Charged pion multiplicity as function of neutron over proton ratio for collisions of different Sn isotopes at $E_{lab}=0.27 A$GeV. The UrQMD-CMF calculations (red and black lines) are compared to experimental results shown as symbols \cite{SpiRIT:2021gtq}.}
    \label{fig:pi_mult_sp}
\end{figure}

Figure \ref{fig:pi_mult_sp} shows the average event multiplicity of charged pions as function of the neutron over proton fraction for $E_{\mathrm{lab}}=0.27 A$ GeV. The UrQMD-CMF simulations are again shown for two scenarios of the iso-spin per baryon fraction, $I_3/B=0$ and $I_3/B=-0.1$. The effect on pion production at this low beam energy is larger than for the higher beam energies, but still relatively small and mostly for the negatively charged pion. In general we can see, that the model describes the $\pi^-$ production rather well for all three systems while the largest deviation is observed for the system with the largest iso-spin asymmetry. The deviation from the simulation is in the opposite direction of the effect of the iso-spin dependence of the potentials. A larger effect is observed for the $\pi^+$ which seems suppressed in the data compared to the simulations. This is similar to the FOPI results where at the lowest beam energies the measured $\pi^+$ multiplicity was overestimated by the model.

\begin{figure} [t]
    \centering
    \includegraphics[width=\columnwidth]{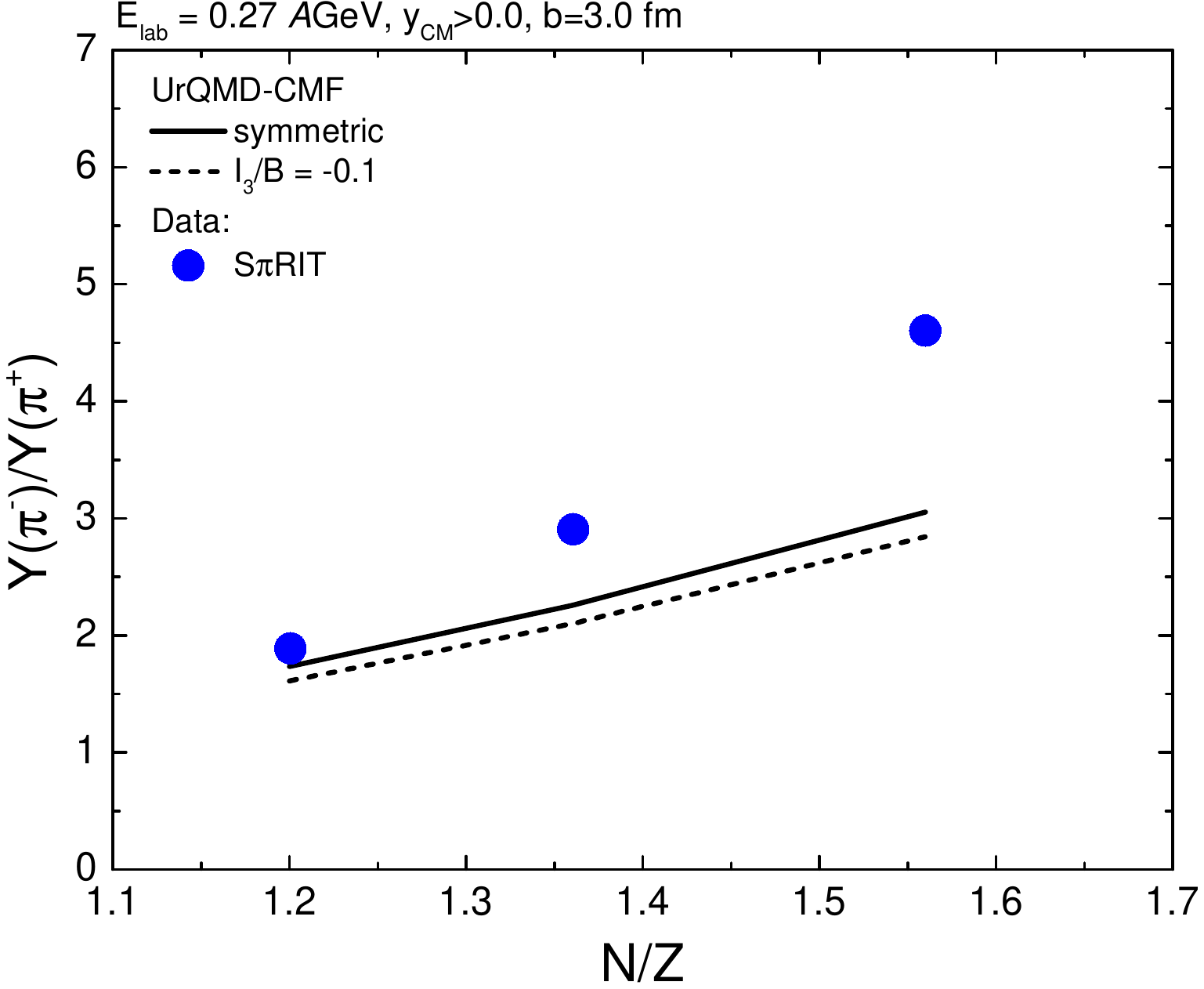}
    \caption{Charged pion ratio as function of neutron over proton ratio for collisions of different Sn isotopes at $E_{lab}=0.27 A$GeV. Data are taken from \cite{SpiRIT:2021gtq}.}
    \label{fig:pi_ratio_sp}
\end{figure}

This effect can be seen more clearly in figure \ref{fig:pi_ratio_sp} where the ratio of the integrated pion yields is shown as function of the neutron to proton ratio. The two lines show again the effect of the iso-spin asymmetry in the potentials which is very small compared to the overall deviation to the data and also goes into the wrong direction. While for systems with small iso-spin asymmetry, the UrQMD-CMF model gives a reasonable description, the discrepancy rises with increasing asymmetry. As we have seen before this is mainly an effect of reduced $\pi^+$ production in the data as compared to the model calculations.

\begin{figure} [t]
    \centering
    \includegraphics[width=\columnwidth]{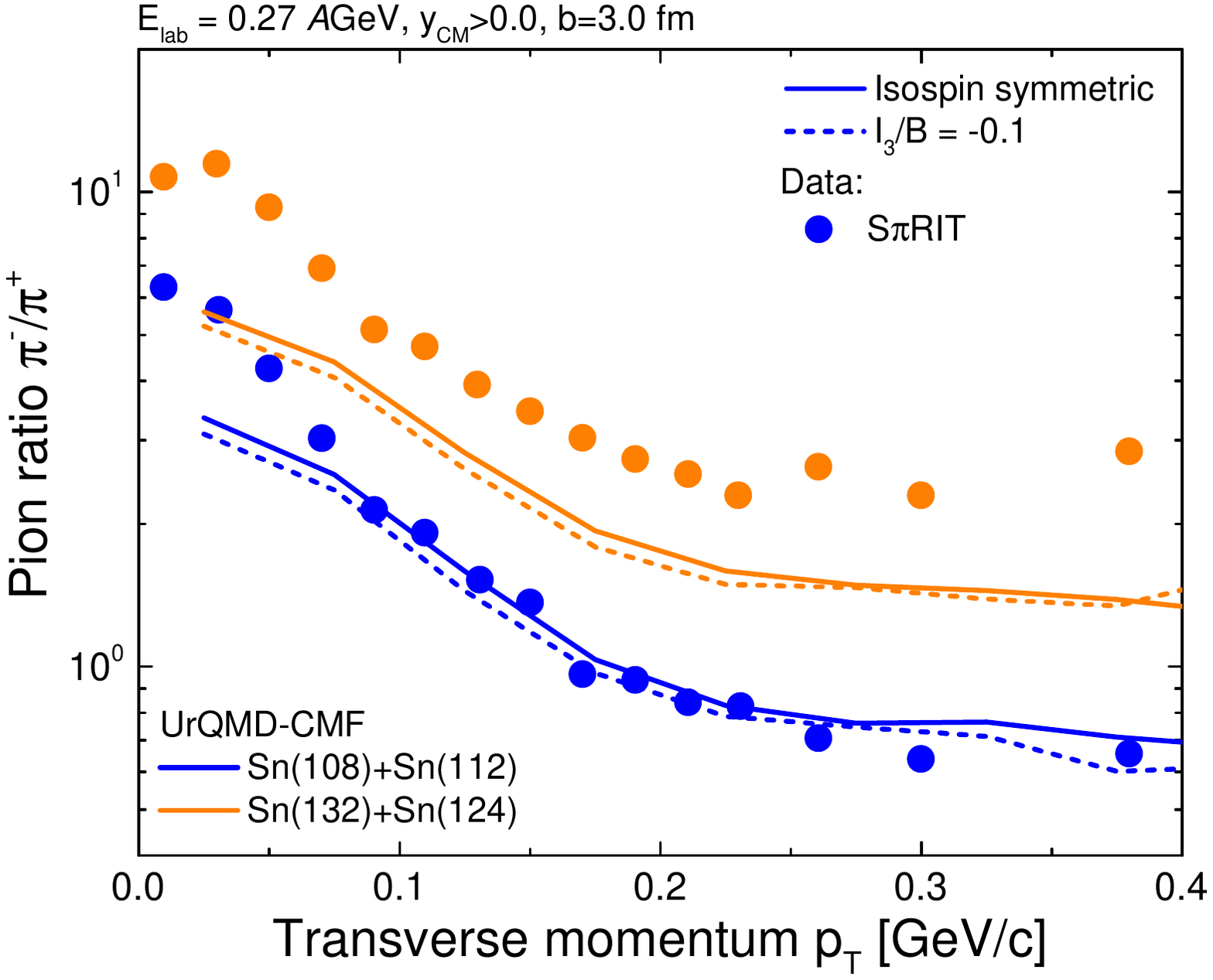}
    \caption{Charged pion ratio as function of transverse momentum for collisions of different Sn isotopes at $E_{lab}=0.27 A$GeV. Data are from \cite{SpiRIT:2021gtq}.}
    \label{fig:pi_pt_sp}
\end{figure}

Finally, looking at the momentum dependence of the charged pion ratio, for the systems measured by S$\pi$RIT, we can again see the effect of the Coulomb interaction as shown in figure \ref{fig:pi_pt_sp}. Negatively charges pions are attracted by the systems bulk (mainly positively charged protons and neutrons) and slowed down while positively charged pions are feeling a repulsion and get accelerated. For the system with the smallest iso-spin asymmetry (blue symbols and lines in figure \ref{fig:pi_pt_sp}) the data can be well described by the UrQMD-CMF simulation, except for very low $p_T$. The system with higher iso-spin asymmetry is qualitatively described, i.e. the shape is correct, but there is a fixed shift in the ratio. This shift simply comes from the difference in multiplicity. Again, the deviation at low $p_T$ is obvious. The two line styles (solid and dashed) represent the two different iso-spin fraction scenarios from the CMF model. As before, the effect is very small and points to the wrong direction. 

\section{Discussion}

We have made an extensive comparison of the CMF model with constraints for the high density equation of state of QCD at varying iso-spin fractions. Our study is based on the effective chiral mean field model which allows us to calculate the equation of state as well as nuclear interactions for different physical systems like neutron stars and heavy ion collisions in a consistent way.

Fixing nuclear matter properties such as the binding energy, saturation density, nuclear incompressibility, symmetry energy and slope of the symmetry energy, the CMF is able to provide a good description of neutron star mass-radius relations and the momentum dependence of the proton single particle potential in nuclear matter. The finite temperature EoS at vanishing net baryon density is consistent with lattice QCD calculations.

Using this parameter set, we make simulations of heavy ion reactions at various beam energies and different iso-spin fractions. It is shown that the data on elliptic flow from the HADES and FOPI experiments can be described reasonably, although the description of the FOPI data is worse than for the HADES data. The reason is not clear as both experiments measure the same systems at almost identical beam energies. This implies that the systematic uncertainties when comparing these two experiments are probably larger than anticipated. Future comparisons of the two experiments at lower beam energies can help shed more light on the uncertainty in the flow measurements which translates directly to an uncertainty in the EoS.

Furthermore, it was shown that the charged pion production near threshold only shows a very weak dependence on the iso-spin asymmetry. UrQMD-CMF provides a very good description for systems that are almost iso-spin symmetric but overestimates $\pi^+$ production for more asymmetric systems. The reason for this is still unclear but unlikely related to the symmetry energy. A possible reason for this effect could be the initialization of the Fermi moment of protons and neutron in UrQMD which are independent of the initial proton fraction, where in reality a more asymmetric system should have smaller Fermi momenta for protons. The transverse dependence of the ratio of charged pions at these low beam energies is mostly sensitive to the Coulomb potential and if its effect are included for mesons the model provides a good description of the data, except for very low transverse momenta. Again, this effect is not well understood and is unlikely related to the symmetry energy as it goes in the wrong direction. Again, different Fermi momenta for protons and neutrons could make a difference for collisions near the pion threshold. Other works have explored the effects of meson-baryon potentials (see e.g. \cite{Cozma:2016qej}) and one may also speculate about possible effects from the fact that pions are bosons and thus should feel an attraction for very small relative momenta. This is, however, something that we plan to explored in a future work.

\section*{acknowledgments}

The computing resources where provided by the Green Cube at GSI and the Center for Scientific Computing of the GU Frankfurt and the Goethe--HLR. The publication is funded by the Open Access Publishing Fund of GSI Helmholtzzentrum fuer Schwerionenforschung.

\bibliography{ref_bib}

\end{document}